\documentstyle[12pt,epsfig,a4]{article} 
\newcommand{\vs}{\vspace{-0.25cm}}
\begin{document} 

\begin{center}
{\Large{\bf Radiative corrections to pion Compton scattering}}
\bigskip

N. Kaiser and J.M. Friedrich\\
\medskip
{\small Physik-Department, Technische Universit\"{a}t M\"{u}nchen,
    D-85747 Garching, Germany}
\end{center}
\medskip
\begin{abstract}
We calculate the one-photon loop radiative corrections to charged pion Compton
scattering, $\pi^- \gamma \to \pi^- \gamma $. Ultraviolet and infrared
divergencies are both treated in dimensional regularization. Analytical
expressions for the ${\cal O}(\alpha)$ corrections to the invariant Compton
scattering amplitudes, $A(s,u)$ and $B(s,u)$, are presented for 11 classes of 
contributing one-loop diagrams. Infrared finiteness of the virtual radiative 
corrections is achieved (in the standard way) by including soft photon 
radiation below an energy cut-off $\lambda$, and its relation to the 
experimental detection threshold is discussed. We find that the
radiative corrections are maximal in backward directions, reaching e.g. 
$-2.4\%$ for a center-of-mass energy of $\sqrt{s}=4m_\pi$ and $\lambda=5\,$MeV. 
Furthermore, we extend our calculation of the radiative corrections by 
including the leading pion structure effect (at low energies) in form of its 
electric and magnetic polarizability difference, $\alpha_\pi - \beta_\pi
\simeq  6\cdot 10^{-4}\,$fm$^3$. We find that this structure effect does not 
change the relative size and angular dependence of the radiative corrections
to pion Compton scattering. Our results are particularly relevant for
analyzing the COMPASS experiment at CERN which aims at measuring the pion
electric and magnetic polarizabilities with high statistics using the
Primakoff effect.    
\end{abstract}

\bigskip
\bigskip

\section{Introduction and summary}
Pion Compton scattering, $\pi^- \gamma \to\pi^- \gamma$, allows one to extract 
the electric and magnetic polarizabilities of the (charged) pion. In a 
classical picture these polarizabilities characterize the deformation response 
(i.e. induced dipole moments) of a composite system in external electric and 
magnetic fields. In the proper quantum field theoretical formulation the 
electric and magnetic polarizabilities, $\alpha_\pi$ and $\beta_\pi$, are
defined as expansion coefficients of the Compton scattering amplitudes at 
threshold.  However, since pion targets are not directly available, real pion
Compton scattering has been approached using different artifices, such as 
high-energy pion-nucleus bremsstrahlung $ \pi^- Z\to\pi^- Z\gamma$, radiative
pion photoproduction off the proton $\gamma p \to \gamma \pi^+ n$, and the 
crossed channel two-photon reaction $\gamma \gamma \to \pi^+\pi^-$.     

From the theoretical side there is an extraordinary interest in a precise
(experimental) determination of the pion polarizabilities. Within the
framework of current algebra \cite{terentev} it has been shown (long ago) that 
the polarizability difference $\alpha_\pi-\beta_\pi$ of the charged pion is 
directly related to the axial-vector-to-vector form factor ratio $h_A/h_V
\simeq 0.44$ measured in the radiative pion decay $\pi^+\to e^+ \nu_e \gamma$ 
\cite{frlez}. At leading (nontrivial) order the result of chiral perturbation 
theory \cite{doho}, $\alpha_\pi- \beta_\pi =\alpha(\bar l_6-\bar l_5)/(24 \pi^2 
f_\pi^2 m_\pi) +{\cal O}(m_\pi)$, is of course the  same after identifying the 
combination of low-energy constants as $\bar l_6-\bar l_5 = 6h_A/h_V +{\cal O}
(m^2_\pi)$. Recently, the systematic corrections to this current algebra result 
have been worked out in refs.\cite{buergi,gasser} by performing a full 
two-loop  calculation of pion Compton scattering in chiral perturbation
theory. The outcome of that extensive analysis is that altogether the higher 
order corrections are rather small and the value $\alpha_\pi- \beta_\pi=(5.7\pm
1.0)\cdot 10^{-4}\,$fm$^3$ \cite{gasser} for the pion polarizability
difference stands now as a firm prediction of the (chiral-invariant) theory. 
The non-vanishing value $\alpha_\pi +\beta_\pi=(0.16\pm 0.1) \cdot 10^{-4}\,$fm$^3$
for the pion polarizability sum (obtained also at two-loop order) is 
presumably too small to cause an observable effect in low-energy pion Compton  
scattering. 

However, the chiral prediction $\alpha_\pi- \beta_\pi=(5.7\pm 1.0)\cdot 
10^{-4}\,$fm$^3$ is in conflict with the existing experimental determinations 
of $\alpha_\pi- \beta_\pi=(15.6\pm 7.8) \cdot 10^{-4}\,$fm$^3$ from Serpukhov 
\cite{serpukhov} and $\alpha_\pi-\beta_\pi =(11.6\pm 3.4)\cdot 10^{-4}\,$fm$^3$ 
from Mainz \cite{mainz}, which amount to values more than twice as large. 
These existing experimental determinations of $\alpha_\pi-\beta_\pi$ certainly
raise doubt as to their correctness since they violate the chiral low-energy 
theorem notably by a factor 2. In that contradictory situation it is promising 
that the ongoing COMPASS \cite{compass} experiment at CERN aims at measuring 
the pion polarizabilities with high statistics using the Primakoff effect. The 
scattering of high-energy negative pions in the Coulomb field of a heavy 
nucleus gives (in the region of sufficiently small photon virtualities) access 
to cross sections for $\pi^- \gamma$ reactions. As an alternative, one
could also directly analyze the bremsstrahlung process $ \pi^- Z\to\pi^-
Z\gamma$, omitting the whole kinematical extrapolation from virtual to real 
photons.  In the appendix we will write down the corresponding fivefold 
differential  cross section $d^5\sigma/d\omega d\Omega_\gamma d\Omega_\pi$
including Born terms, the pion polarizability difference
$\alpha_\pi-\beta_\pi$, and an (equally) important pion-loop correction.

In any case, the effects of the pion's low-energy structure on (real or 
virtual) Compton scattering observables turn out to be relatively small. For 
center-of-mass energies $\sqrt{s} < 4m_\pi$ (i.e. sufficiently below the 
prominent $\rho(770)$-resonance) the differential cross sections $d\sigma/
d\Omega_{\rm cm}$ in backward directions are reduced at most by about $11\%$ in 
comparison to the ones of a structureless pion \cite{picross}. Therefore, 
a precise knowledge of the pure QED radiative corrections to pion Compton
scattering is indispensable if one wants to extract the pion polarizabilities 
from the data with good accuracy. In certain kinematical regions the effects
from the pion's low-energy  structure and the pure QED radiative corrections 
may become of comparable size. Although calculations of radiative corrections 
are abundant in the literature, we are not aware of a detailed and utilizable 
exposition of the radiative corrections to scalar (spin-0) boson 
Compton scattering. The case of spin-1/2 electron Compton scattering has been 
treated already in the early days of quantum electrodynamics by Brown and 
Feynman \cite{feyn}. The radiative corrections to the Thomson limit, valid 
close to threshold, have been reported in ref.\cite{corin}.  In the work of 
Akhundov et al. \cite{akhundov} the one-photon loop diagrams to virtual pion
Compton scattering have been considered, but no accessible sources to the 
corresponding analytical expressions (which are necessary for an 
implementation into data analyses) are given. Since those previous results
were mainly presented in numerical form it is difficult to implement them
independently into future data analyses. As a consequence of that deficit 
the radiative corrections to pion Compton scattering are sometimes merely 
adapted from the known ones for myons by simply replacing the mass: $m_\mu \to 
m_\pi$. Of course, such a substitutional procedure mistreats profound 
differences in the couplings of photons to scalar spin-0 bosons and 
spin-1/2 fermions. 

The purpose of the present paper is to fill this gap and to present a detailed 
calculation of the one-photon loop radiative corrections to pion Compton 
scattering, $\pi^- \gamma \to \pi^- \gamma$. We give closed-form analytical 
expressions for the ${\cal O}(\alpha)$ corrections to the two invariant 
amplitudes $A(s,u)$ and  $B(s,u)$ as they emerge from 11 classes of 
contributing one-loop diagrams. Ultraviolet and infrared divergencies are both
treated by the method of dimensional regularization which ensures gauge
invariance at every step. While the ultraviolet divergences can be absorbed in
renormalization constants, the cancellation of infrared divergencies requires 
the inclusion of soft photon radiation below an experimental detection 
threshold $\lambda$. The total finite radiative correction depends then 
logarithmically on the small energy resolution scale $\lambda$. We find that 
the radiative corrections become maximal in backward directions, reaching 
about $-2.4\%$ at a center-of-mass energy of $\sqrt{s} = 4m_\pi$ for  $\lambda
=5\,$MeV. With such a size and kinematical signature the radiative corrections 
are not negligible in comparison to the effects from the pion 
polarizabilities, which also show up preferentially in the backward 
directions. Furthermore, we include in our calculation of the radiative 
corrections also the leading pion structure effect through a two-photon 
contact-vertex proportional to the pion polarizability difference $\alpha_\pi
-\beta_\pi \simeq 6 \cdot 10^{-4}\,$fm$^3$. We find that these structure
effects do not modify the relative size and angular dependence of the
radiative  corrections. Our results can be utilized for analyzing the COMPASS 
experiment at CERN.

\section{Pion Compton scattering amplitude}
We start out with defining the invariant amplitudes for the (real) pion 
Compton scattering process: $\pi^-(p_1)+\gamma(k_1,\epsilon_1)\to \pi^-(p_2)+
\gamma(k_2,\epsilon_2)$. It is advantageous to work in the center-of-mass frame
and to choose (in this frame) the Coulomb gauge $\epsilon_{1,2}^0=0$ for the 
(transverse) photon polarization vectors. The corresponding T-matrix reads 
then:
\begin{equation} T_{\pi \gamma} = 8\pi \alpha \bigg\{ - \vec \epsilon_1
\cdot\vec \epsilon_2^{\,\,*} \, A(s,u) +  \vec \epsilon_1 \cdot \vec k_2\,  \vec 
\epsilon_2^{\,\,*} \cdot\vec k_1 \, {2\over t} \Big[  A(s,u)+ B(s,u) \Big] \bigg\}
\,, \end{equation}
with $\alpha=e^2/4\pi= 1/137.036$, and $s=(p_1+k_1)^2\geq m_\pi^2$ and $t=(k_1-k_2
)^2 \leq0$ the two independent Mandelstam variables. The inequalities refer to 
the physical region of pion Compton scattering and the third (dependent)
Mandelstam variable $u = (p_1-k_2)^2 \leq m_\pi^2$ is given by $u=2m_\pi^2-s-t$. As
indicated in eq.(1) we view the two (dimensionless) invariant amplitudes, 
$A(s,u)$ and $B(s,u)$, as  functions of $s$ and $u$. The (three) tree diagrams 
of  scalar quantum electrodynamics lead to the following contributions: 
\begin{equation} A(s,u)^{(\rm tree)}=1\,,\qquad B(s,u)^{(\rm tree)}={s-m_\pi^2  \over 
u-m_\pi^2} \,. \end{equation}
Note that the contribution of the $s$-channel pole diagram vanishes, since the 
couplings of the initial and final state photon are both equal to zero in 
Coulomb gauge, $\epsilon_1\cdot (2p_1+k_1) = 0=\epsilon_2\cdot(2p_2+k_2)$, in 
the center-of-mass frame. Performing the sums over transverse photon 
polarizations and applying flux and appropriate two-body phase space factors 
the differential cross section at tree level reads: 
\begin{equation} {d \sigma^{(\rm pt)}\over d\Omega_{\rm cm}}= {\alpha^2[s^2
(1+z)^2+ m_\pi^4(1-z)^2]  \over s[s(1+z)+m_\pi^2(1-z)]^2}\,,\end{equation}
where $t = (s-m_\pi^2)^2(z-1)/2s$ has been expressed in terms of the cosine of
the cms scattering angle $z=  \cos \theta_{\rm cm}= \hat k_1 \cdot \hat k_2$. 
When including the one-photon loop radiative corrections into the invariant
amplitudes, $A(s,u)$ and $B(s,u)$, the differential cross section gets 
modified by an additive interference term (see eq.(3) in ref.\cite{picross}): 
\begin{eqnarray} {d \sigma_{\rm int} \over d\Omega_{\rm cm}} &=& {\alpha^2 
\over s[s(1+z)+m_\pi^2(1-z)] } \bigg\{ 2m_\pi^2(1-z)\, {\rm Re}\,A(s,u)^{(
\gamma-\rm  loop)}\\\nonumber &&+(1+z)[m_\pi^2(1-z)-s(1+z)] \, {\rm Re}\,B(s,u)
^{(\gamma-\rm loop)} \bigg\} \,,\end{eqnarray}
of order $\alpha^3$. Its ratio to the point-like cross section in eq.(3) 
defines the (virtual part of the) radiative correction factor $\delta_{\rm virt} 
\sim \alpha$ as a function of the center-of-mass energy $\sqrt{s}$ and $z=\cos
\theta_{\rm cm}$.  

Before we turn to the explicit evaluation of the one-photon loop diagrams in 
the next section we add several remarks about regularization and 
renormalization. We use the method of dimensional regularization to treat both 
ultraviolet and infrared divergencies (where the latter are caused by the 
masslessness of the photon). The method consists in calculating loop integrals 
in $d$ spacetime dimensions and expanding the results around $d=4$. Divergent 
pieces of one-loop integrals generically show up in form of the composite 
constant:
\begin{equation} \xi = {1\over d-4} +{1\over2} (\gamma_E-\ln 4\pi) +\ln{m_\pi
\over \mu}\,, \end{equation}
containing a simple pole at $d=4$. In addition, $\gamma_E = 0.5772\dots$ is the 
Euler-Mascheroni number and $\mu$ an arbitrary mass scale introduced in 
dimensional regularization in order to keep the mass dimension of the loop 
integrals independent of $d$. Ultraviolet (UV) and infrared (IR) divergencies 
are distinguished by the feature of whether the condition for convergence of 
the $d$-dimensional integral is $d<4$ or $d>4$. We discriminate them in the 
notation by putting appropriate subscripts, i.e. $\xi_{UV}$ and $\xi_{IR}$. In 
the actual calculation an infrared divergent term originates typically from a 
Feynman parameter integral of the form: $\int_0^1 dy \, y^{d-5}  = 1/(d-4)_{IR}$. 

Since scalar quantum electrodynamics is a multiplicatively renormalizable 
field theory, all (ultraviolet) divergencies can be absorbed in renormalization
constants $Z_j$ (see appendix A in ref.\cite{radiat}). These establish the 
(formal) relations between (renormalized) physical quantities ($m_\pi, e$,
etc.) and the bare ones.\footnote{The mass counterterm which cancels on the 
mass-shell the pion self-energy correction from the photon-loop is $\delta 
m_\pi^2 =\alpha m_\pi^2(6 \xi_{UV}-7)/4\pi$.} In the present context, we want 
to remind the reader only about the cancellation between self-energy 
corrections and vertex corrections for the electric charge $e$. For that 
purpose, consider the coupling of a zero-momentum  photon with polarization 
vector $\epsilon$ to an on-shell pion of momentum $p$. The tree level coupling 
$-2ie\, \epsilon \cdot p$ supplemented by the one-photon loop corrections to
it takes the following form:  
\begin{equation} -2i e \,\epsilon \cdot p \, \bigg\{ 1 + {\alpha \over 4\pi} 
\Big[ (4\xi_{IR}-4\xi_{UV})+(6 \xi_{UV}-7) +(7-4\xi_{IR}-2\xi_{UV}) \Big] 
\bigg\} \,, \end{equation} 
where the terms in round brackets correspond to the three diagrams shown in 
Fig.\,1 in that order (including for the first two their horizontally
reflected partners). The same cancellation mechanism ($Z_1=Z_2$) is at work in 
spinor electrodynamics \cite{radiat}, where the central 
one-photon loop diagram in Fig.\,1 is absent, and the balance in the square
bracket of eq.(6) reads: $(4\xi_{IR}+2\xi_{UV}-4)+(4-4\xi_{IR}-2\xi_{UV})=0$. The 
effect of the pionic vacuum polarization (on the photon propagator) is 
canceled by the counterterm $Z_3 -1 =\alpha \xi_{UV}/6\pi$, corresponding the 
$1/4$ of the value in spinor electrodynamics \cite{radiat}.     

\begin{figure}
\begin{center}
\includegraphics[scale=1.1,clip]{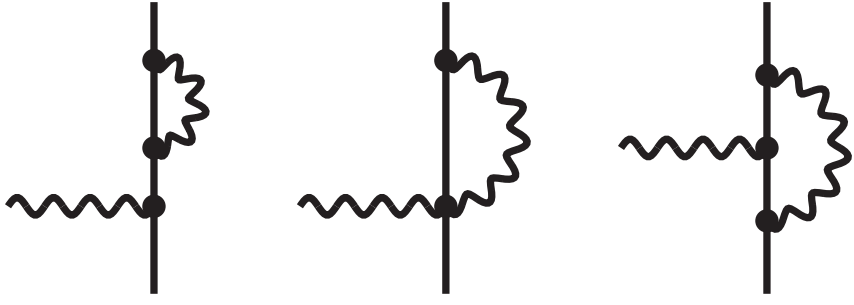}
\end{center}
\vspace{-.8cm}
\caption{One-photon loop diagrams contributing to charge renormalization in
scalar QED. Additional diagrams obtained by a horizontal reflection are not 
shown. A self-energy correction on an external line (left graph) brings $1/2$ 
of the pion wavefunction renormalization. In total the effects sum to
zero.} 
\end{figure}
\section{Evaluation of one-photon loop diagrams}
In this section, we present analytical results for the radiative corrections
of order $\alpha$ to the invariant amplitudes $A(s,u)$ and $B(s,u)$ of pion
Compton scattering. Our convenient choice of Coulomb gauge (in the 
center-of-mass frame) eliminates still all $s$-channel pole diagrams since for 
these either the initial state or final state photon coupling vanishes 
identically. Dropping also the diagrams with a tadpole-type self-energy 
insertion (the loop consisting of a single virtual photon line, which is set 
to zero in dimensional regularization) we are left with the 20 one-photon loop 
diagrams shown in Fig.\,2. For the purpose of an ordering scheme they have
been divided into classes I-XI. Due to an increasing number of internal pion 
propagators their evaluation rises in complexity. In order to simplify all
calculations, we exploit gauge invariance and employ the Feynman gauge where 
the photon propagator is just proportional to the Minkowski metric tensor 
$g^{\mu\nu}$. Moreover, for a concise presentation of the one-loop amplitudes, 
$A(s,u)^{(\gamma-\rm loop)}$ and $B(s,u)^{(\gamma-\rm loop)}$,
it is most helpful to work with the dimensionless kinematical variables $\hat 
s= s/m_\pi^2 \geq 1$, $\hat t =t/m_\pi^2 \leq 0$ and $\hat u =u/m_\pi^2 \leq1$, 
which obey the relation  $\hat s+\hat t +\hat u=2$.

\begin{figure}
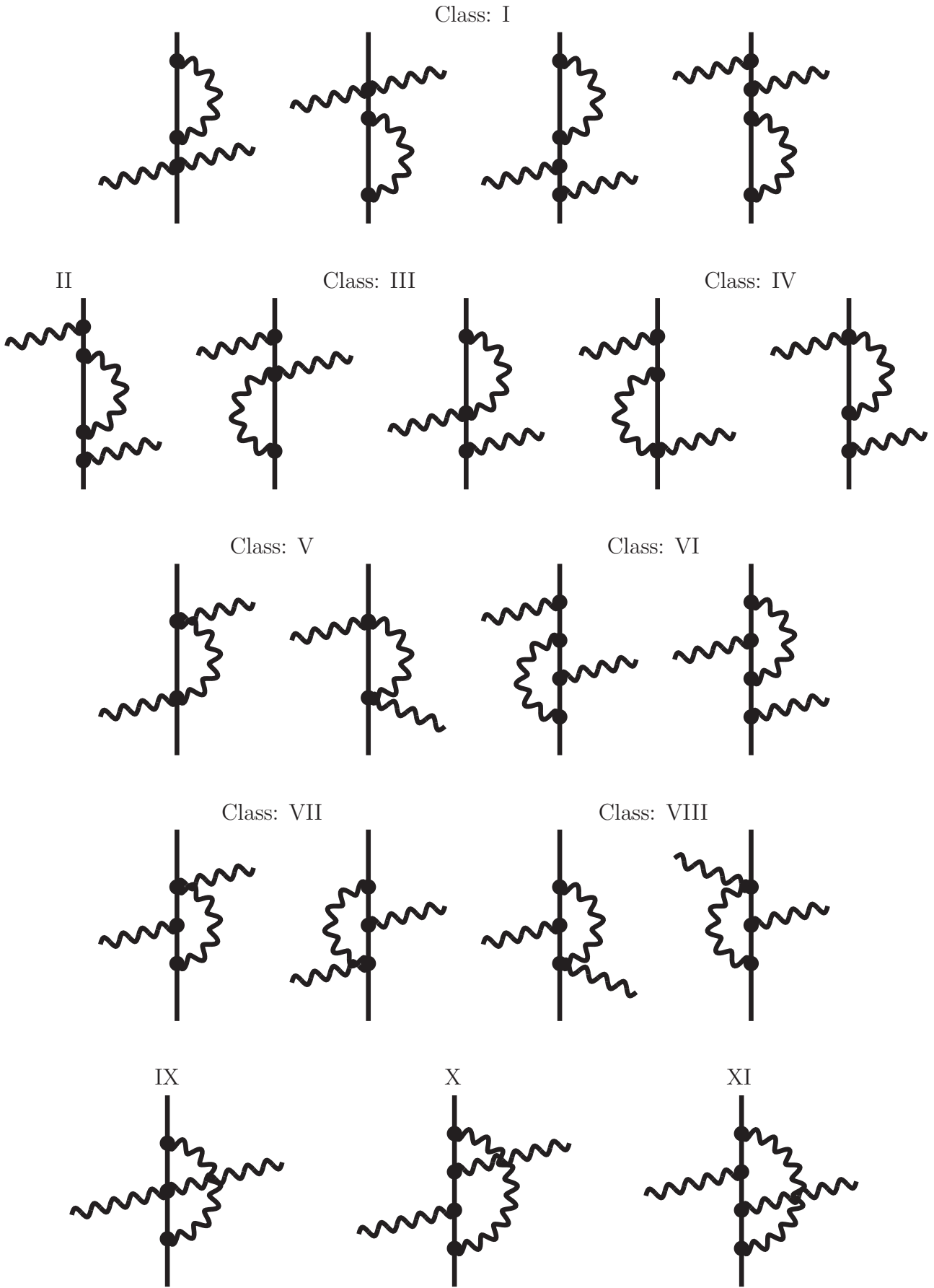

\begin{center}
\includegraphics[scale=1.,clip]{radfig2.epsi}

\bigskip
\bigskip

\includegraphics[scale=1.,clip]{radfig3.epsi}

\bigskip
\bigskip

\includegraphics[scale=1.,clip]{radfig4.epsi}

\bigskip
\bigskip

\includegraphics[scale=1.,clip]{radfig5.epsi}

\bigskip
\bigskip

\includegraphics[scale=1.,clip]{radfig6.epsi}
\end{center}
\vspace{-.3cm}
\caption{One-photon loop diagrams for radiative corrections to pion Compton
scattering.}
\end{figure}

We can now enumerate the analytical expressions for the 11 classes of 
contributing one-photon loop diagrams. For notational simplicity we drop the
arguments $s$ and $u$ of the invariant amplitudes $A(s,u)$ and  $B(s,u)$, as
well as a superscript indicating the class of diagrams.

\medskip

\noindent
\underline{Class\,I:} For this class of diagrams the tree level amplitudes in 
eq.(2) get multiplied  by the wavefunction renormalization factor of the
pion:\footnote{In our scheme of book keeping each self-energy correction on an 
external (on-shell) pion-line yields for the scattering amplitude one half of 
the pion wavefunction renormalization factor $Z_2-1=\alpha(\xi_{IR}-\xi_{UV})/
\pi$.}  
\begin{equation}  A= {\alpha \over \pi} \Big(\xi_{IR}-\xi_{UV}\Big) \,,  
\qquad  B= {\alpha \over \pi}\, {\hat s-1 \over \hat u-1}  \Big(\xi_{IR}-
\xi_{UV} \Big) \,. \end{equation} 

\medskip
\noindent
\underline{Class\,II:} This $u$-channel pole diagram involves the 
once-subtracted (off-shell) self-energy of the pion:
\begin{equation}  A= 0\,, \qquad B= {\alpha \over 2\pi}\, {\hat t \over \hat 
u-1} \bigg[ 2\xi_{UV}-2 +{\hat u+1\over \hat u}\ln(1-\hat u)\bigg]  \,.
\end{equation} 

\medskip
\noindent
\underline{Class\,III:} These $u$-channel pole diagrams generate a constant
vertex correction factor:
\begin{equation}  A= 0\,, \qquad B= {\alpha \over 2\pi}\, {\hat t \over \hat 
u-1} \bigg[ -3\xi_{UV} +{7\over 2}\bigg]  \,. \end{equation}

\medskip
\noindent
\underline{Class\,IV:} These $u$-channel pole diagrams  generate a $\hat
u$-dependent vertex correction factor:
\begin{equation}  A= 0\,, \qquad B= {\alpha \over 2\pi}\, {\hat t \over \hat 
u-1} \bigg[-3\xi_{UV} +3+{1\over 2\hat u} +{1-\hat u\over 2\hat u^2}(3\hat u+1)
\ln(1-\hat u)\bigg]\,. \end{equation}

\medskip
\noindent
\underline{Class\,V:} The sum of photon rescatterings in the $s$- and 
$u$-channel gives:
\begin{equation}  A= - B= {\alpha \over 2\pi}\bigg[4\xi_{UV} -4+{\hat s-1\over
\hat s} \ln(1- \hat s)+{\hat u-1\over \hat u} \ln(1-\hat u)\bigg]\,. 
\end{equation}

\medskip
\noindent
\underline{Class\,VI:} These $u$-channel pole diagrams  generate again a $\hat
u$-dependent vertex correction factor:
\begin{equation}  A= 0\,, \qquad B= {\alpha \over 2\pi}\, {\hat t \over \hat   
 u-1} \bigg[ 2\xi_{UV} -{5\over 2}-{1\over 2\hat u}-{\hat u^2+6 \hat u+1\over
 2\hat   u^2} \ln(1-\hat u)\bigg]\,,\end{equation} 

\medskip
\noindent
\underline{Class\,VII:} These irreducible $s$-channel diagrams give:
\begin{equation}  A= - B= {\alpha \over 2\pi}\bigg\{-\xi_{UV} +{3\over 2}-{\hat
    s+1 \over 2\hat s}\ln(1-\hat s)+ {1\over \hat s-1} \bigg[{\pi^2 \over 6}
-{\rm Li}_2(\hat s)\bigg] \bigg\}\,, \end{equation}
where  Li$_2(x) = \sum_{n=1}^\infty n^{-2} x^n =  x \int_1^\infty
dy [y(y- x)]^{-1} \ln y$ denotes the conventional dilogarithmic function.

\medskip
\noindent
\underline{Class\,VIII:} These irreducible $u$-channel diagrams give:
\begin{eqnarray} && A={\alpha \over 2\pi}\bigg\{-\xi_{UV} +{3\over 2}-{\hat u
+1 \over 2\hat u}\ln(1-\hat u)+ {1\over \hat u-1}\bigg[{\pi^2 \over 6}-{\rm 
Li}_2(\hat u) \bigg] \bigg\}\,, \\ && B=-A+ {\alpha \over 2\pi}\, {\hat t 
\over \hat u-1}\bigg\{{3\hat u-1 \over 2\hat u^2}\ln(1-\hat u)-{\hat u+1\over 
2\hat u} +{1\over \hat  u-1} \bigg[{\rm Li}_2(\hat u)-{\pi^2 \over 6} 
\bigg] \bigg\}\,.\end{eqnarray}

\medskip
\noindent
\underline{Class\,IX}: The $t$-dependent vertex correction to the two-photon 
contact coupling reads:     
\begin{eqnarray} A=-B\!\!\!&=&\!\!\!{\alpha\over 2\pi}\bigg\{-\xi_{UV} +1+
\sqrt{4-\hat t}\, L(t) + {\hat t-2 \over \sqrt{\hat t^2-4\hat t}}\bigg[4 
\xi_{IR} \sqrt{-\hat t}\,L(t)  \nonumber \\ && \qquad +{\rm  
Li}_2(w)-{\rm  Li}_2(1-w) +{1\over 2} \ln^2 w -{1\over 2}  \ln^2(1-w) \bigg] 
\bigg\}\,. \end{eqnarray}
Here, we have introduced the frequently occurring logarithmic loop function: 
\begin{equation} L(t) = {m_\pi\over \sqrt{-t}} \ln{\sqrt{4m_\pi^2 -t}+
\sqrt{-t} \over 2 m_\pi} \,, \end{equation} 
and the auxiliary variable $w$ appearing in eq.(16) is determined by the 
relation $1-2w  = \sqrt{-t/(4m_\pi^2-t)}$.

\medskip
\noindent
\underline{Class\,X:} The $s$-channel box diagram gives:
\begin{eqnarray} A\!\!\!&=&\!\!\!{\alpha\over 2\pi}\bigg\{ {1\over 2}\xi_{UV}-
{3\over 4}+{\hat s+1 \over 2\hat s}\ln(1-\hat s)+ {1\over \hat s-1} \bigg[{\rm
  Li}_2(\hat s)-{\pi^2 \over 6}\bigg]-{1\over 2}\sqrt{4-\hat t}\, L(t) 
\nonumber \\ && -L^2(t)+\int_1^\infty \!\!{dx \over x-\hat s}\, {2-\hat t\over 
(x-1)^2+\hat  t x} \Big[(x+1)\ln x+2(1-x)\sqrt{4-\hat t} \, L(t)\Big]\bigg\}
\,, \end{eqnarray} 
\begin{eqnarray} B\!\!\!&=&\!\!\!-A+ {\alpha \over 2\pi}\bigg\{ {1\over 4} 
-L^2(t)+\int_1^\infty\!\!{dx \over x-\hat s}\, {2-\hat t\over [(x-1)^2+\hat t 
 x]^2} \bigg[\hat t x (x+1) \ln x  \nonumber \\ && + [(x-1)^2+\hat t x](x-1)
+{2(x-1) \over \sqrt{4-\hat t}} [\hat t^2x -6\hat t x -2(x-1)^2]L(t) 
\bigg]\bigg\}\,. \end{eqnarray}
The loop integrals involving four propagators are written here in terms of 
their spectral function representations (employing the Kramers-Kronig 
dispersion relation), where the $x$-dependent integrands stand for the 
corresponding imaginary parts. For the numerical evaluation of the relevant 
real parts of $A$ and $B$ one has to treat $\int_1^\infty  dx/(x-
\hat  s)\dots $ as a principal value integral. It can be conveniently 
decomposed into a sum of two  nonsingular integrals by the following formula: 
\begin{equation} -\!\!\!\!\!\!\int_1^\infty \!\!dx \, {f(x) \over x-\hat s} = 
\int_1^{2\hat s-1}\!\!dx\, {f(x)- f(\hat s)\over x-\hat s} + \int_{2\hat s-1}
^\infty \!\!dx \, {f(x)\over x-\hat s} \,. \end{equation} 

\medskip
\noindent
\underline{Class\,XI:} The $u$-channel box diagram gives:
\begin{eqnarray} A\!\!\!&=&\!\!\!{\alpha\over 2\pi}\bigg\{ {1\over 2}\xi_{UV}-
{3\over 4}+{\hat u+1 \over 2\hat u}\ln(1-\hat u)+ {1\over\hat u-1} \bigg[{\rm
  Li}_2(\hat u)-{\pi^2 \over 6}\bigg]-{1\over 2}\sqrt{4-\hat t}\, L(t)\nonumber
\\ && -L^2(t)+\int_1^\infty \!\!{dx \over x-\hat u}\,{2-\hat t\over  (x-1)^2+
\hat t x} \Big[(x+1)\ln x+2(1-x)\sqrt{4-\hat t}\, L(t)\Big]  \bigg\}\,, 
\end{eqnarray}

\begin{eqnarray} B\!\!\!&=&\!\!\!-A+ {\alpha \over 2\pi}\Bigg\{ {\hat t(\hat 
u+1)\over  2\hat u(\hat u-1)}+ {\hat t(1-\hat u-2\hat u^2) \over 2\hat u^2
(\hat u-1)}\ln(1-\hat u) +{\hat t\over (\hat u-1)^2} \bigg[{\pi^2  \over 6}
-{\rm  Li}_2(\hat u) \bigg]\nonumber \\ && +{1\over 4} -L^2(t)+ {\hat t^2-2
\hat t \over  (1-\hat u)  \sqrt{\hat t^2-4\hat t}}\bigg[4\Big(\xi_{IR}+\ln(1
-\hat u)\Big) \sqrt{-\hat t}\,L(t) +{\rm  Li}_2(w)
\nonumber \\ &&  -{\rm Li}_2(1-w) +{1\over 2} \ln^2 w -{1\over 2} \ln^2(1-w) 
+{\rm Li}_2(h_-) - {\rm Li}_2(h_+) \bigg]+\int_1^\infty\!\!{dx \over x-\hat u}
\, \nonumber \\ && \times  {2-\hat t\over [(x-1)^2+\hat t x]^2} \bigg\{\hat t
(x+1)(2-\hat t-x) \ln x+[(x-1)^2+\hat t x]\Big(x+\hat t+{\hat  t\over  
x}-1\Big) \nonumber \\ && +{2L(t)\over \sqrt{4-\hat t}}\Big[\hat t^3(x+1)+
\hat t^2(3x^2-3x-4) +2\hat t(x^3-4x^2+2x+1)-2(x-1)^3\Big] \bigg\} \Bigg\}\,, 
\end{eqnarray}
where the arguments $h_{\pm }$ of the dilogarithms Li$_2(h_\pm)$ are determined 
by the relation $2m_\pi^2h_{\pm }= t \pm  \sqrt{t^2-4 m_\pi^2 t}$.   

Several checks of our calculation can be performed now. First, one verifies 
that the ultraviolet divergent terms proportional to $\xi_{UV}$ cancel in the 
total sums for $A$ and $B$ separately (using the relation $\hat s + \hat u + 
\hat t = 2$ in the latter case). Secondly, crossing symmetry implies for our 
chosen set of independent amplitudes that $A(s,u)$ and $(u-m_\pi^2)(s-m_\pi^2)^{-1}
B(s,u)$ are symmetric functions under the exchange of the variables, $s 
\leftrightarrow u$. This property follows e.g. by relating our set of
amplitudes, $A(s,u)$ and $B(s,u)$ defined via eq.(1), to the (manifestly) 
crossing-symmetric functions introduced in refs.\cite{buergi,gasser}. While 
the symmetry condition $A(s,u) = A(u,s)$ becomes evident after summing up all 
contributions, the one for $B(s,u)$ provides a highly nontrivial check 
between analytical terms and terms represented by dispersion integrals in
eqs.(18-22). In its consequence, crossing symmetry interrelates the whole set 
of one-loop diagrams in Fig.\,2, which as such is not manifestly
crossing-symmetric due to the  particular choice of Coulomb gauge.

\section{Infrared finiteness}
In the next step we have to consider the infrared divergencies. Inspection of 
eqs.(7,16,22) reveals that the (summed) infrared divergent terms proportional 
to $\xi_{IR}$ appear in the same ratio $A(s,u):B(s,u)$ as the (tree level) Born 
terms written in eq.(2). As a consequence of that, the infrared divergent 
virtual (loop) corrections multiply the point-like cross section
$d\sigma^{(\rm  pt)}/d \Omega_{\rm cm}$ by a  factor: 
\begin{equation} \delta_{\rm virt}^{(\rm IR)}= {2 \alpha \over \pi} \Bigg[ 1+
{2 \hat t -4 \over \sqrt{4- \hat t}} L(t) \Bigg] \, \xi_{IR} \,. \end{equation}
The unphysical infrared divergence $\xi_{IR}$ gets canceled at the level of 
the cross section by the contributions of soft photon bremsstrahlung. In its 
final effect, the (single) soft photon radiation  multiplies the tree level 
cross section $d\sigma^{(\rm pt)}/d\Omega_{\rm cm}$  by a (universal) factor 
\cite{feyn,radiat}:
\begin{equation} \delta_{\rm soft}= \alpha\, \mu^{4-d}\!\!\int\limits_{|\vec 
l\,|<\lambda} \!\!{d^{d-1}l  \over (2\pi)^{d-2}\, l_0} \bigg\{ {2m_\pi^2 - t 
\over p_1 \cdot l \, p_2 \cdot l} - {m_\pi^2 \over (p_1 \cdot l)^2} - {m_\pi^2 
\over (p_2 \cdot l)^2} \bigg\} \,, \end{equation}
which depends on a small energy  resolution $\lambda$. Working out this
momentum space integral by the method of dimensional regularization (with 
$d>4$) one finds that the infrared divergent correction factor $\delta_{\rm
  virt}^{(\rm IR)} \sim \xi_{IR}$ gets removed and the following finite radiative 
correction factor remains:
\begin{equation}\delta_{\rm real}= {\alpha \over \pi}\Bigg\{ \bigg[ 2+ {4 \hat 
t -8 \over\sqrt{4- \hat t}} L(t) \bigg] \ln{m_\pi \over 2\lambda} + { \hat s+1 
\over \hat s-1} \ln \hat s + \int_0^{1/2}\!\! dy \,{(\hat s+1)( \hat t-2)\over 
W [1-  \hat t y(1-y)] } \ln{ \hat s+1+W \over  \hat s+1-W }\Bigg\} \,,   
\end{equation}
with the abbreviation $W = \sqrt{(\hat s-1)^2 + 4 \hat s \hat t y(1-y)}$.
One can easily convince oneself that the radiative correction factor $\delta_{
\rm real}$ vanishes in forward direction $t=0$. We note that the terms beyond 
the ones proportional to $\ln(m_\pi /2\lambda)$ are specific for the evaluation 
of the soft photon correction factor $\delta_{\rm soft}$ in eq.(24) in the 
center-of-mass frame. As it is written in eq.(25), $\delta_{\rm real}$ refers to 
an (idealized) experimental situation where all undetected soft photon 
radiation  would fill in momentum space a small sphere of radius $\lambda$ in 
the center-of-mass frame. In a real experiment this region will be of
different shape with no sharp boundaries due the detector efficiencies etc. 
Such additional (experiment specific) radiative corrections can be accounted
for and calculated by integrating the fivefold differential cross section for
$\pi^-\gamma \to \pi^- \gamma \gamma$ \cite{picross} over the appropriate
region in  phase space. By construction this region excludes the infrared 
singular domain $|\vec l\,| <\lambda$  and thus leads to a finite result.  

We are now in the position to present numerical results for the radiative
corrections to pion Compton scattering. The complete radiative correction
factor is $\delta_{\rm real}$ written in eq.(25) plus all finite loop amplitudes 
$A$ and $B$ inserted into the interference cross section $d\sigma_{\rm  int}/d
\Omega_{\rm cm}$ and divided by the tree level cross section $d\sigma_{\rm  pt}/d
\Omega_{\rm cm}$. Fig.\,3 shows in percent the total radiative correction factor 
for three selected center-of-mass energies $\sqrt{s}=(2,3,4)m_\pi$ as a 
function of $z= \cos \theta_{\rm cm}$. The detection threshold for soft photons 
has been set to the value $\lambda = 5\,$MeV. One observes that the radiative 
corrections become maximal in backward directions $z\simeq -1$, reaching 
values up to $-2.4\%$ for $\sqrt{s} = 4m_\pi$. With such an angular dependence 
the pure QED radiative corrections have the same kinematical signature as the 
effects from the pion's low-energy structure (i.e. pion polarizability 
difference $\alpha_\pi - \beta_\pi$ and pion-loop correction of chiral
perturbation theory \cite{picross}). In magnitude they are suppressed by about 
a factor $5-10$. We agree qualitatively with the findings of 
ref.\cite{akhundov} that the radiative corrections simulate effects of (fake) 
pion polarizabilities $\alpha_\pi^{(rc)} = - \beta_\pi^{(rc)} \simeq 3 \cdot 
10^{-5}\,$fm$^{3}$ corresponding in magnitude to about $10\%$ of the prediction 
of chiral perturbation theory \cite{gasser}. A proper inclusion of radiative 
corrections is therefore essential if one wants to extract the pion 
polarizabilities from Compton scattering data with good accuracy. We also note 
that when increasing the energy resolution scale $\lambda$ the magnitude of 
the radiative correction goes down. For example, with $\lambda = 10\,$MeV or 
$20\,$MeV the before-mentioned $-2.4\%$ change to $-1.8\%$ or $-1.2\%$.  

Finally, it is worth to mention that as one approaches (at fixed scattering 
angle $z = \cos \theta_{\rm  cm}$) the Compton threshold, $\sqrt{s}\to m_\pi$, all 
radiative correction effects go away \cite{corin}. In this long-wavelength 
limit only the pure Thomson amplitude $T_{\pi  \gamma} = -8\pi \alpha \,\vec 
\epsilon_1 \cdot\vec \epsilon_2^{\,\,*}$ survives. This non-renormalization
theorem is  strictly fulfilled by the one-loop amplitudes $A(s,u)$ and 
$B(s,u)$ written in eqs.(7-22). In the same way, a direct computation at zero 
photon momentum $k_1=k_2=0$ gives for the factor multiplying the Thomson
amplitude: 
\begin{eqnarray} &&1+ {\alpha \over 4\pi} \Big[ (4\xi_{IR}-4\xi_{UV})^{\rm (I)}+ 
(8\xi_{UV}-8)^{\rm (V)}+ (1-2\xi_{UV})^{\rm (VII)}+(1-2\xi_{UV})^{\rm (VIII)} \nonumber \\ && 
\qquad \qquad +(4-4\xi_{IR}-2\xi_{UV})^{\rm (IX)}+(\xi_{UV}+1)^{\rm
  (X)}+(\xi_{UV}+1)^{\rm  (XI)} \Big] \,, \end{eqnarray}     
where we have indicated the (classes of) contributing diagrams.

\begin{figure}
\begin{center}
\includegraphics[scale=.55,clip]{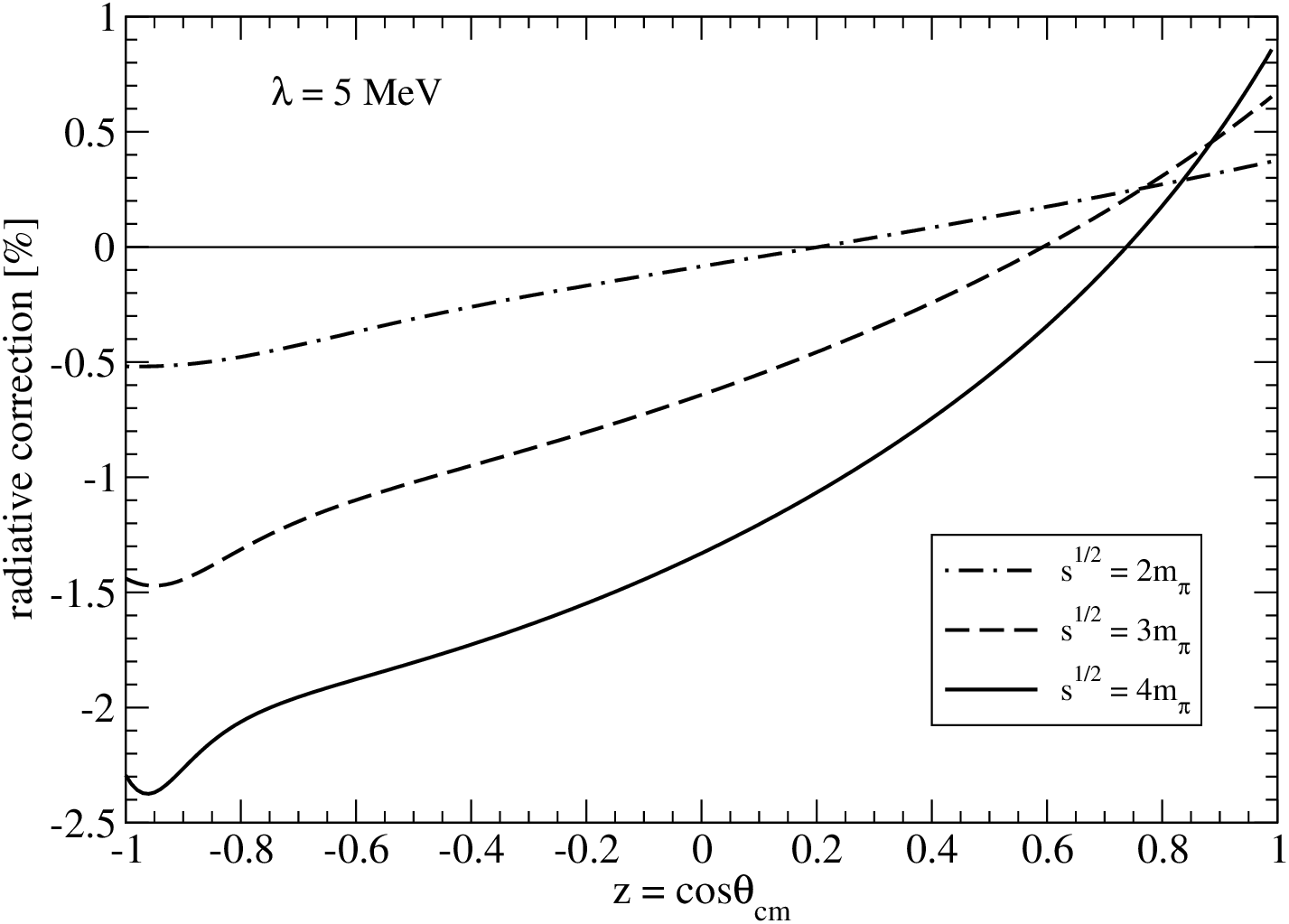}
\end{center}
\vspace{-.8cm}
\caption{Radiative corrections for pion Compton scattering $\pi^- \gamma \to
\pi^- \gamma$. The pion is treated as a structureless spin-0 boson.}
\end{figure}

\section{Radiative corrections including pion polarizabilities}
So far we have treated in our calculation of the radiative corrections the
pion as a structureless spin-0 boson. Such an approximation is valid only at
very low (photon) energies. The leading pion structure relevant for Compton 
scattering is given by the difference of its electric and magnetic 
polarizability $\alpha_\pi- \beta_\pi$. In an effective field theory approach
it is easily accounted for by a new two-photon contact-vertex proportional to 
the squared  electromagnetic field strength tensor, $F_{\mu\nu}F^{\mu\nu} = 2(\vec
B^2- \vec E^2)$. The S-matrix insertion
following from this (higher-order) gauge-invariant effective
$\pi\pi\gamma\gamma$-vertex reads: 
\begin{equation} 8\pi i \beta_\pi m_\pi \Big( k_1 \cdot k_2 \, \epsilon_1  \cdot 
\epsilon^*_2 -\epsilon_1\cdot k_2\,\epsilon^*_2\cdot k_1\Big)\,,\end{equation} 
where one photon $(k_1, \epsilon_1)$ is ingoing and the other one $(k_2, 
\epsilon_2)$ outgoing. At tree level the polarizability vertex eq.(27) gives 
rise to the contributions $A(s,u)^{(\rm pola)} = - \beta_\pi m_\pi t/2\alpha$ and 
$B(s,u)^{(\rm pola)}=0$ to the invariant Compton scattering amplitudes. In
order to prevent any misunderstandings we stress that when writing (merely) 
$\beta_\pi m_\pi$ for the coupling strength in eq.(27), an electric and a 
magnetic pion polarizability, equal in magnitude and opposite in sign
$\alpha_\pi = -\beta_\pi$, are always both included.

We can reinterpret the one-photon loop diagrams in Fig.\,2 now in such a way 
that the two-photon contact-vertex represents the polarizability vertex 
proportional to $\beta_\pi m_\pi$. In the case of class\,V we have to treat two 
cases, namely either the upper or lower contact-vertex. Going through the 
classes I, III-IX and reevaluating the loop diagrams with the S-matrix 
insertion from the polarizability vertex, we find the following contributions 
to the invariant amplitudes $A$ and $B$. 
 
\medskip
\noindent  
\underline{Class\,I:}
\begin{equation} A={\beta_\pi m_\pi \, t \over 2\pi} \Big(\xi_{UV}-\xi_{IR} \Big)
\,,  \qquad  B= 0 \,, \end{equation} 
\medskip
\noindent  
\underline{Class\,V:}
\begin{eqnarray} && A= {\beta_\pi m_\pi^3\over 2\pi}\bigg[-\xi_{UV}\,\hat t+\hat t 
+1-{1\over 2 \hat s}-{1\over 2 \hat u}+{( \hat s-1)^3\over 2\hat s^2} \ln(1-
\hat s)+{( \hat u-1)^3\over 2\hat u^2} \ln(1- \hat u) \bigg] \,, \\ &&  B=-A+ 
{\beta_\pi m_\pi \, t \over 2\pi}\bigg[ -\xi_{UV}+1-{1\over 2 \hat u} -{(\hat  u-
1)^2\over 2\hat u^2} \ln(1- \hat u) \bigg] \,, \end{eqnarray}
\medskip
\noindent  
\underline{Class\,VII:}
\begin{eqnarray} && A= {\beta_\pi m_\pi^3 \over 2\pi}(\hat u-1)\bigg\{\xi_{UV}  -{3
\over 2}+{\hat s+1 \over 2\hat s}\ln(1-\hat s)+ {1\over \hat s-1} \bigg[
{\rm Li}_2(\hat s)-{\pi^2 \over 6}\bigg] \bigg\}\,, \\ && B = {\hat s -1\over 
\hat u-1} \,A \,, \end{eqnarray}

\medskip
\noindent  
\underline{Class\,VIII:}
\begin{eqnarray} &&A={\beta_\pi m_\pi^3 \over 2\pi}(\hat s-1)\bigg\{\xi_{UV} -{3
\over 2}+{\hat u+1 \over 2\hat u}\ln(1-\hat u)+ {1\over \hat u-1}\bigg[ {\rm 
Li}_2(\hat u)-{\pi^2 \over 6} \bigg] \bigg\}\,, \\ && B=-A+ {\beta_\pi m_\pi
\,t \over 2\pi} \,  {\hat s -1\over \hat u-1}\bigg\{{\hat u+1\over  2\hat u}+
{1-3\hat u \over 2\hat u^2}\ln(1-\hat u)+{1\over \hat  u-1} \bigg[{\pi^2 \over 
6} -{\rm Li}_2(\hat u)\bigg] \bigg\}\,,\end{eqnarray}
\medskip
\noindent  
\underline{Class IX:}
\begin{eqnarray} && A={\beta_\pi m_\pi\,t \over 4\pi}\bigg\{\xi_{UV} -1- \sqrt{4
-\hat t}\, L(t) + {2-\hat t \over \sqrt{\hat t^2-4\hat t}}\bigg[4 \xi_{IR} 
\sqrt{-\hat t}\,L(t) \nonumber \\ && \qquad  +{\rm  Li}_2(w)-{\rm  Li}_2(1-w) 
+{1\over 2} \ln^2 w -{1\over 2}  \ln^2(1-w) \bigg] \bigg\}\,, \qquad B = 0\,. 
\end{eqnarray}
The contributions from classes III and IV (multiplicative vertex correction
factors) vanish due to some special features of the S-matrix insertion 
eq.(27). 

One verifies that the ultraviolet divergent terms proportional to $\xi_{UV}$ 
drop out in the total sum for $B(s,u)$, but not for $A(s,u)$. In the latter
case the $t\,\xi_{UV}$-term can be canceled by interpreting the coupling 
constant in eq.(27) as a bare one and splitting it as $\beta_\pi^{(\rm bare)} =
\beta_\pi (1 - \alpha\, \xi_{UV}/2\pi)$ into the physical coupling constant
$\beta_\pi$ and a counterterm piece.\footnote{Note that the polarizability 
vertex involves a coupling strength $\beta_\pi m_\pi$ of negative mass 
dimension and therefore on general grounds it will generate a 
non-renormalizable quantum field theory.} One also convinces oneself easily
that the crossing-symmetry relations are fulfilled for the total sums of 
$A(s,u)$ and $B(s,u)$. The infrared divergent terms proportional to $\xi_{IR}$ 
showing up in eqs.(28,35) are again canceled by the soft photon bremsstrahlung
contributions. As a result, one has to multiply the partial cross section
linear in the polarizability:  
\begin{equation} {d \sigma^{(\rm pola)}\over d\Omega_{\rm cm}}= {\alpha \beta_\pi 
m_\pi^3 (s- m_\pi^2)^2(1-z)^2 \over 2s^2[s(1+z)+m_\pi^2(1-z)]}\,,\end{equation}
with the radiative correction factor $\delta_{\rm real}$ written down in eq.(25).

\begin{figure}
\begin{center}
\includegraphics[scale=.55,clip]{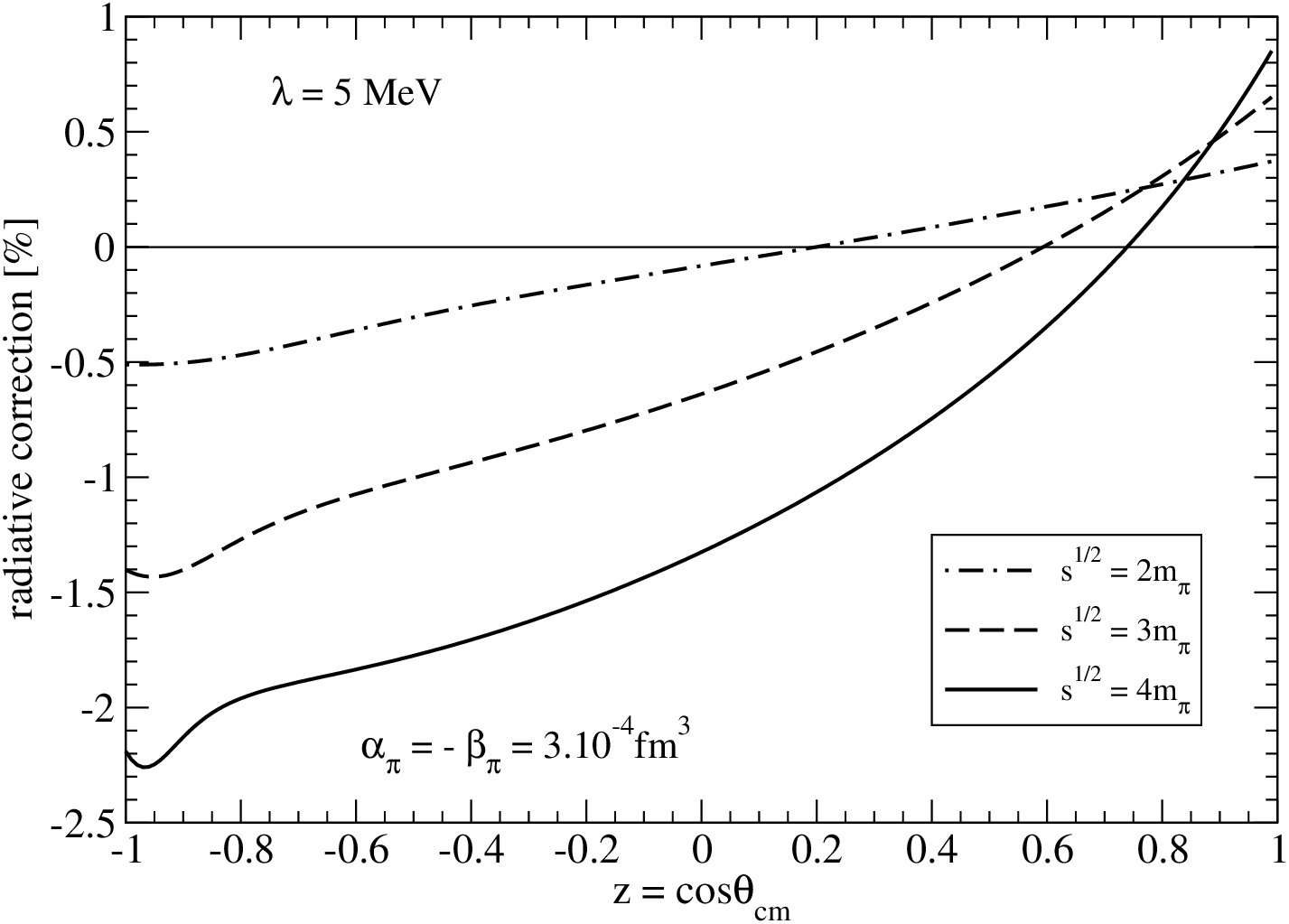}
\end{center}
\vspace{-.8cm}
\caption{Radiative corrections for pion Compton scattering $\pi^- \gamma \to
\pi^- \gamma$. The pion's low-energy structure is included through a
polarizability difference $\alpha_\pi- \beta_\pi \simeq 6 \cdot 10^{-4}\,$fm$^{3}$.}
\end{figure}

In Fig.\,4 we show again the radiative correction factor for center-of-mass
energies $\sqrt{s} =(2,3,4) m_\pi$ as a function of $z= \cos \theta_{\rm cm}$ 
with a pion polarizability difference of $\alpha_\pi- \beta_\pi \simeq 6 \cdot 
10^{-4}\,$fm$^{3}$ included. The reference cross section is now the sum 
${d \sigma^{(\rm  pt)}/d\Omega_{\rm    cm}}+ {d \sigma^{(\rm  pola)}/  d\Omega_{\rm cm}}$ of 
point-like and polarizability-improved ($\sim \beta_\pi$) cross section. The 
radiative correction is of course for each part of relative order $\alpha$. 
The detection threshold of soft photons has been kept at the value $\lambda = 
5\,$MeV. In comparison to Fig.\,3 one observes almost no changes for the three
curves in Fig.\,4. Only some slight upward shifts of the maximal values near $
z \simeq -1$ are visible. We can therefore conclude that the size and angular 
dependence of the radiative corrections to pion Compton scattering remain
practically unchanged when including the leading structure effect in form of 
the pion polarizability difference $\alpha_\pi- \beta_\pi \simeq 6 \cdot
10^{-4} \,$fm$^{3}$. Of course, there is yet the pion-loop correction of chiral 
perturbation theory \cite{buergi,gasser}. In the cross sections it works 
partly against the polarizability effects \cite{picross}. It is expected that 
the combined radiative pion-photon two-loop corrections (emerging from a vast
number of two-loop diagrams) will follow the same trend.        

In summary, we have presented a detailed (analytical) calculation of radiative 
corrections to pion Compton scattering which can be utilized for data analyses.

\section*{Appendix: Cross section for pion-nucleus bremsstrahlung}
It is common practice to extract from Primakoff events in high-energy
pion-nucleus bremsstrah- lung a cross section for pion Compton scattering via 
the equivalent photon method \cite{pomer}. This involves a kinematical
extrapolation from (time-like polarized) virtual photons to (space-like 
transverse) real photons. While the procedure becomes exact for asymptotically
small photon virtualities $q^2 \to0$, it introduced uncertainties into the 
analysis of real experiments, where only some lower bounds can be reached. 
Moreover, at the very low photon-virtualities $q^2$ shielding effects of the 
nuclear charge by inner-shell atomic electrons comes into play as a further 
complication. The issue of Coulomb-strong interaction interference in 
pion-nucleus bremsstrahlung has been  studied recently in ref.\cite{faeldt}. 
According to this work it is advantageous to restrict oneself to momentum 
transfers $|\vec q\,| \leq 32 \,$MeV $\simeq m_\pi/4$, since there the ratio 
between 
polarizability and Born contributions remains essentially the same as for real
pion Compton scattering.     

We want to point out here, that for the purpose of determining the pion 
polarizabilities (i.e. the dominant difference $\alpha_\pi- \beta_\pi$) one 
can omit the whole extrapolation procedure and directly analyze the pion
bremsstrahlung process as it is measured in the laboratory frame. The relevant
ingredients to the cross section (at sufficiently low momentum transfers 
$|\vec q\,|$ where the one-photon exchange dominates) are summarized in this
appendix.  
      
Consider the pion bremsstrahlung process, $\pi^-(\vec p_1) + Z_{\rm rest} \to 
\pi^-(\vec p_2)+ \gamma(\vec k, \vec \epsilon\,) + Z_{\rm recoil}$, where a
virtual photon transfers the (small) momentum $\vec q =\vec p_2+\vec k-\vec p_1$
from the nucleus (of charge $Z$) to the pion. The extremely small recoil energy
$-q_0 = \vec q\,^2/2M_{\rm nucl} \approx 0 $ of the nucleus can be neglected. The 
fivefold differential cross section in the laboratory  frame reads 
\cite{itzykson}: 
\begin{equation} {d^5 \sigma \over d\omega d\Omega_\gamma d\Omega_\pi} = {Z^2
    \alpha^3 \omega p_2 \over \pi^2 p_1 q^4} \,H\,, \end{equation}
with $q = |\vec q\,|$ and $H$ the squared amplitude summed over the final 
state  photon polarizations $(\vec \epsilon \cdot \vec k = 0)$: 
\begin{equation} H = (\tilde A p_1\sin \theta_1)^2+(\tilde B  p_2\sin\theta_2)^2 
+ 2\tilde A \tilde B  p_1  p_2\sin \theta_1 \sin \theta_2 \cos \phi\,.
\end{equation} 
This expression has been derived from a T-matrix for the virtual Compton 
scattering process, $\pi^-\gamma^* \to \pi^-\gamma$, of the form $T = -8\pi
\alpha (\tilde A \,\vec \epsilon\,^* \cdot \vec p_1+ \tilde B\,\vec\epsilon\,^* 
\cdot \vec p_2) $ with $\tilde A $ and $\tilde B$ two (real) amplitudes. 
$\theta_1$ and $\theta_2$ are the angles between the momentum vectors $\vec k$ 
and $\vec p_1$ and $\vec p_2$, respectively. In addition, $\phi$ is the angle 
between the two planes spanned by $(\vec k,\vec p_1)$ and $(\vec k, \vec p_2)$. 

The (three) tree diagrams of scalar quantum electrodynamics lead to the
following contributions to the amplitudes $\tilde A $ and $\tilde B$: 
\begin{equation}\tilde A^{(\rm tree)} = {E_2 \over \omega (E_1 - p_1 \cos \theta_1)}
  \,, \qquad \tilde B^{(\rm tree)} = {-E_1 \over \omega (E_2 - p_2 \cos \theta_2)}\,,
\end{equation}
with $E_1 = \sqrt{p_1^2 + m_\pi^2}$ the initial energy of the pion, $E_2= E_1 
-\omega$ the final energy of the pion and $p_2 = \sqrt{p_1^2-2 E_1\omega+
\omega^2}$. The photon energy is denoted by $\omega$. Note that the diagram
involving the two-photon contact-vertex of scalar QED vanishes here, due to the
orthogonality of the time-like virtual photon and the space-like real photon 
polarizations, $\epsilon_0 =0$. The contributions from the pion electric and 
magnetic polarizabilities (taking $\alpha_\pi =- \beta_\pi$ as a good
approximation) read:  
\begin{equation} \tilde A^{(\rm pola)} = -\tilde  B^{(\rm pola)} = { \beta_\pi m_\pi 
\omega  \over \alpha} \,. \end{equation}
It follows directly from the S-matrix insertion in eq.(27) evaluated in the 
present kinematical situation, $\pi^-\gamma^* \to \pi^-\gamma$.  At the same 
order (in the small momentum expansion) there are the pion-loop corrections of 
chiral perturbation theory \cite{unkmeir}. On the one hand side, they 
introduce the charge form  factor of the pion:
\begin{equation} F_\pi(q^2) = 1 - {q^2 \over 6} \langle r^2_\pi\rangle +  {1\over 
12 \pi^2 f_\pi^2} \bigg\{ -m_\pi^2 -{q^2 \over 3} +{(4m_\pi^2+q^2)^{3/2} \over  4q} 
\tilde L(q) \bigg\} \,, \end{equation}
with $\langle r^2_\pi\rangle= (0.45\pm0.01)$\,fm$^2$ \cite{pdg} the mean square 
charge radius  of the pion, $f_\pi = 92.4\,$MeV the pion decay constant, and
the logarithmic loop function:
\begin{equation}\tilde L(q) = \ln {\sqrt{4m_\pi^2 +q^2}+ q \over 2m_\pi} \,.
\end{equation}

\begin{figure}
\begin{center}
\includegraphics[scale=1.1,clip]{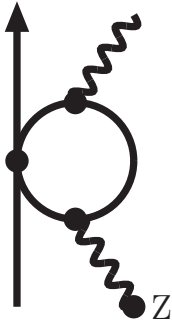}
\end{center}
\vspace{-.3cm}
\caption{Loop diagram for virtual pion Compton scattering.} 
\end{figure}
As a result, the tree level Born terms, $\tilde A^{(\rm tree)}$ and $\tilde
B^{(\rm tree)}$ in eq.(39), have then to be multiplied by the pion charge form 
factor $F_\pi(q^2)$, because a virtual photon (with $q^2 \ne 0$) coupling to 
the pion  can resolve this part of structure of the pion. Moreover and 
specific for (virtual) Compton scattering, there is the pion-loop diagram 
shown in Fig.\,5. It describes photon scattering off the ''pion-cloud around 
the pion'' and leads to the following contribution \cite{unkmeir}:     
\begin{eqnarray} \tilde A^{(\rm loop)}= -\tilde B^{(\rm loop)} &=& {\omega \tau^2 \over 
(4\pi f_\pi)^2 (\tau^2-q^2)^2} \bigg\{\tau^2-q^2 + 4m_\pi^2 \Big[ \tilde L^2(q)- 
\tilde L^2(\tau)  \Big]\nonumber \\ && + 2 q \sqrt{4m_\pi^2 +q^2}\,\tilde L(q) 
-{2q^2 \over \tau} \sqrt{4m_\pi^2  +\tau^2}\, \tilde L(\tau) \bigg\} \,.  
\end{eqnarray}
Note that this irreducible loop contribution depends on both, the photon 
virtuality $q^2$ and the squared invariant momentum transfer $-\tau^2$ between 
the initial and final state pion, with $\tau^2 = \vec q\cdot ( \vec q-2 \vec k 
\,) = (\vec p_1-\vec p_2)^2 -\omega^2$ where $\vec p_1\cdot  \vec p_2 = p_1 p_2 (
\cos \theta_1 \cos\theta_2 + \sin \theta_1  \sin \theta_2 \cos\phi)$. In the
cross section, the pion-loop correction in eq.(43) compensates in part the 
effects from the pion polarizabilities \cite{picross}. Therefore, it is 
important to  include it in the analysis of pion bremsstrahlung data, if one 
wants to extract reliably the pion polarizability difference $\alpha_\pi -
\beta_\pi$. 

Of course, there are yet the one-photon loop radiative corrections to  
virtual pion Compton scattering $\pi^-\gamma^* \to \pi^-\gamma$ (see Fig.\,2).
These will be presented (in the same style as done here for real pion Compton
scattering) in a forthcoming publication \cite{next}.


\begin{thebibliography}{99}
\bibitem{terentev} M.V. Terentev, {\it Sov. J. Nucl. Phys.} {\bf 16}, 87 
(1973).\vs
\bibitem{frlez} E. Frlez et al., {\it Phys. Rev. Lett.} {\bf 93}, 181804 
(2004); hep-ex/0606023.\vs
\bibitem{doho} J.F. Donoghue and B.R. Holstein,  {\it Phys. Rev.} {\bf D40},
2378 (1989).\vs
\bibitem{buergi} U. B\"urgi,  {\it Phys. Lett.} {\bf B377}, 147 (1996); 
{\it Nucl. Phys.} {\bf B479}, 392 (1996).\vs
\bibitem{gasser} J. Gasser, M.A. Ivanov, and M.E. Sainio, {\it Nucl. Phys.}
{\bf B745}, 84 (2006); and refs. therein.\vs
\bibitem{serpukhov} Y.M. Antipov et al., {\it Phys. Lett.} {\bf B121}, 445 
(1983); {\it Z. Phys.} {\bf C26}, 495 (1985).\vs
\bibitem{mainz} J. Ahrens et al., {\it Eur. Phys. J.} {\bf A23}, 113 (2005).\vs
\bibitem{compass} COMPASS Collaboration: P. Abbon et al., hep-ex/0703049.\vs
\bibitem{picross} N. Kaiser and J.M. Friedrich, {\it Eur. Phys. J.} {\bf A36},
181 (2008).\vs
\bibitem{feyn} L.M. Brown and R.P. Feynman, {\it Phys. Rev.} {\bf 85}, 231 
(1952).\vs
\bibitem{corin} E. Corinaldesi and R. Jost, {\it Helv. Phys. Acta} {\bf 21},
183 (1948).\vs
\bibitem{akhundov} A.A. Akhundov, D.Y. Bardin, and G.V. Mitselmakher, 
{\it Sov. J. Nucl. Phys.} {\bf 37}, 217 (1983); \\ A.A. Akhundov, D.Y. Bardin, 
G.V. Mitselmakher, and A.G. Olshevskii, {\it Sov. J. Nucl. Phys.} {\bf 42},
426 (1985); \\  A.A. Akhundov, S. Gerzon,
S. Kananov, and M.A. Moinester, {\it Z. Phys.} {\bf C66}, 279 (1995).\vs 
\bibitem{radiat} M. Vanderhaeghen et al.,  {\it Phys. Rev.} {\bf C62}, 025501 
(2000); and refs. therein.\vs
\bibitem{pomer} I.Y. Pomeranchuk and I.M. Shmushkevich, {\it Nucl. Phys.} {\bf 
23}, 452 (1961).\vs
\bibitem{faeldt} G. F\"aldt and U. Tengblad, nucl-th/0807.2700; 
nucl-th/0802.0971; {\it Phys. Rev.} {\bf C76}, 064607 (2007).\vs 
\bibitem{itzykson} C. Itzykson and J.B. Zuber, Quantum Field Theory,
McGraw-Hill book company, 1980; chapter 5.2.4.\vs
\bibitem{unkmeir} C. Unkmeir, S. Scherer, A.I. Lvov, and D. Drechsel,  {\it 
Phys. Rev.} {\bf C61}, 034002 (2000).\vs
\bibitem{pdg} Particle Data Group (W.M. Yao et al.), {\it J. Phys.} {\bf G33},
1 (2006).\vs
\bibitem{next} N. Kaiser and J.M. Friedrich, in preparation.\vs
\end{thebibliography}
\end{document}